\documentclass{article}

\usepackage{PRIMEarxiv}

\usepackage[utf8]{inputenc} % allow utf-8 input
\usepackage[T1]{fontenc}    % use 8-bit T1 fonts
\usepackage{hyperref}       % hyperlinks
\usepackage{url}            % simple URL typesetting
\usepackage{booktabs}       % professional-quality tables
\usepackage{amsfonts}       % blackboard math symbols
\usepackage{nicefrac}       % compact symbols for 1/2, etc.
\usepackage{microtype}      % microtypography
\usepackage{lipsum}
\usepackage{fancyhdr}       % header
\usepackage{graphicx}       % graphics
\graphicspath{{media/}}     % organize your images and other figures under media/ folder
\usepackage{epstopdf}
\usepackage{epsfig}
\usepackage{amsmath, bm}
\usepackage{wrapfig}
\usepackage[numbers]{natbib}

\newcommand{\avg}[1][\cdot]{\left\langle {#1} \right\rangle}
\usepackage{xparse}
\NewDocumentCommand{\IP}{ O{\cdot} O{\cdot}}{\left(#1, #2 \right)}

\newcommand{\norm}[1][\cdot]{\left| \left| {#1} \right| \right|}

\DeclareMathOperator*{\argmax}{arg\,max}

\title{Intrinsic phase based proper orthogonal decomposition (IPhaB POD): a method for physically interpretable modes in near-periodic systems
}

\author{
  Akhileshwar Borra, Zoey Flynn, Andres Goza, Theresa Saxton-Fox \\
Department of Aerospace Engineering\\
  University of Illinois Urbana-Champaign\\
  Urbana, IL\\
  \texttt{akhil.borra@gmail.com} \\
}

\begin{document}
\maketitle

\begin{abstract}

Fluid dynamics systems driven by dominant, nearly periodic large-scale dynamics are common across wakes, jets, rotating machinery, and high-speed flows. Traditional decomposition techniques such as proper orthogonal decomposition and dynamic mode decomposition have been used to gain insight into these flows, but can require many modes to represent physical processes. %a single physical process or set of processes.
With the aim of generating modes that intuitively convey the underlying physical mechanisms, we propose an intrinsic phase-based proper orthogonal decomposition (IPhaB POD) method. IPhaB POD creates energetically ranked modes that evolve along a characteristic cycle of a driving near-periodic large scale.  %Related efforts have pursued similar modal decompositions that evolve relative to a large scale, posed in the frequency domain. 
Our proposed formulation is set in the time domain, which is particularly useful in cases where the large-scale dynamics are imperfectly periodic. % and would have to be represented across several frequencies.
We formally derive IPhaB POD within a proper orthogonal decomposition framework, and it inherits the optimal representation inherent to POD. As part of this derivation, a dynamical systems representation of the large scale is utilized, facilitating a definition of phase relative to the large scale within the time domain. An expectation operator and inner product are also constructed relative to this definition of phase in a manner that allows for the various cycles within the data to demonstrate imperfect periodicity. The formulation is tested on two sample problems: a simple, low Reynolds number airfoil wake, and a complex, high-speed pulsating shock wave problem. The resulting modes are shown to better isolate the large-scale dynamics in the first mode than space-only proper orthogonal decomposition, and to highlight meaningful small-scale dynamics in higher modes for the shock flow problem.

\end{abstract}

%\begin{keywords}
%proper orthogonal decomposition (POD), wake flows, shocks, modal analysis
%\end{keywords}

\section{Introduction}

 Dominant, near-periodic large scales are common across many flows, including bluff body wakes, rotating machinery, and shock-laden processes. These systems often exhibit near-periodic large-scale dynamics and coherent structures, as well as turbulent, smaller-scale behaviors that are tethered to the larger-scale flow. Extracting the disparate flow scales and structures and their interactions are critical to understanding, modeling, and controlling these systems. We focus in this article on developing a data-driven decomposition that provides more physically interpretable representations, or ``modes'', of these near-periodic systems (see, e.g., \citet{taira2017modal} for a review of data-driven and operator-based decomposition techniques, which include linear stability analysis, resolvent analysis, proper orthogonal decomposition (POD), and dynamic mode decomposition (DMD)).
 
When used to analyze fluid flows generally, most modal decomposition techniques are not guaranteed to provide insight into instantaneously relevant coherent structures. However, in many cases the resulting modes have been successfully used to investigate the physics of specific scales of motion. Comparing modes between forced and unforced cases can identify the impact of forcing on different flow scales in a jet \citep{kourentis2012uncovering,Cylindershedding1,schmid2011applications}, or can clarify the nature of coherent structures associated with upstream and downstream aero-optic distortions in a turbulent boundary layer \citep{borra2022proper}. Modes associated with specific frequencies can isolate the coherent structures that generate bands of acoustic noise \citep{Airfoil1,freund2009turbulence} and those associated with specific instabilities \citep{pickeringjets, schmidt_towne_rigas_colonius_brès_2018, Huang, PODandDMD}. Decompositions have advanced understanding of the physics of bluff body wakes \citep{kourentis2012uncovering,Cylindershedding1,Cylindershedding2,ping2021dynamic,Airfoil1,Airfoil2}, flows past flexible bodies \citep{schmid2010dynamic,goza2018modal}, cavity flows \citep{Cavity1,Cavity2,Cavity3,Cavity4}, and shear layers \citep{Shear1,Shear2,Shear3}. 
With improved understanding of the behavior of specific flow features, comes an improved ability to control the flow \citep{RowleyandDawson}. Reduced order models constructed through modal analysis techniques have been used to design control algorithms for myriad applications, including re-circulation control in channel flow \citep{Ravindran}, separation control over a backwards facing ramp \citep{Taylor}, a flat plate at large angles of attack \citep{Ahuja}, and the wake of a circular cylinder \citep{Bergmann}. 
 
 %The modes that come out of these analyses are often utilized to physically understand the coherent motions of flows driven by inherent large-scale motions. For example, for the vortex wake of a cylinder, the first few POD modes have been used to compare the structure and energy of that wake between unforced and forced cases \cite{kourentis2012uncovering,Cylindershedding1}. POD can be a useful technique to understand the impact of control because its energy-ranked modes can distill the most energetically significant features of the wake for each case and quantify their energy content, enabling comparisons. 
 
 Despite the utility of these decomposition techniques, we highlight three limitations that have motivated modifications within the community. First, while in some flows, modes derived from decompositions have been found to physically resemble instantaneous structures \citep{saxton2017coherent}, in other flows the modes generated by these decompositions are structurally dissimilar from instantaneous flow structures. To understand coherent structure behavior, researchers have added multiple modes together \citep{Airfoil1} or turned to other techniques entirely such as finite time Lyapunov exponent fields \citep{kourentis2012uncovering} or conditional projection averaging \citep{saxton2022amplitude}.
 % or dynamical mode decomposition \citep{PODandDMD}, to better capture instantaneously similar flow structures. 
Second, modal decompositions also tend to prioritize behaviors that are energetically or dynamically significant statistically, but these modes can miss important rare or extreme events \citep{schmidt2019conditional}. The modes that are identified as most dynamically or energetically significant can vary depending on how long of a time series is considered in multi-scale systems \citep{mendez_balabane_buchlin_2019}. Finally, many of the decomposition techniques are only appropriate for statistically stationary flows.

Modifications to standard variants of modal decomposition techniques have been proposed to address some of these challenges. Non-stationarity has been addressed through space-time variants of POD by \citet{gordeyev2013temporal} for use with control, and by \citet{schmidt2019conditional} to identify mechanisms driving rare loud events in jet noise. Multi-resolution techniques are also naturally suited to non-stationary problems, and multi-resolution DMD, multi-scale POD, and data-driven wavelet techniques have been demonstrated to provide insights into optimal modes over a range of time scales \citep{kutz2016multiresolution,mendez_balabane_buchlin_2019,floryan2021discovering}.

%Implementations of proper orthogonal decomposition for non-stationary flows are demonstrated by \citet{gordeyev2013temporal} for use with control and by \citet{PhasePOD_Fabien} using a phase averaging approach. Finite space-time proper orthogonal decomposition and its conditional form provide insights into localized or rare events \citep{schmidt2019conditional}.  Multi-resolution DMD, multi-scale POD, and data-driven wavelet techniques can provide insights into optimal modes over a range of time scales \citep{kutz2016multiresolution,mendez_balabane_buchlin_2019,floryan2021discovering}. %Each technique generates modes that 

For the systems of focus in this article, which are dominated by near-periodic behaviors, \citet{padovan2020analysis} and \citet{padovan2022analysis} extend the traditional assumption of a stationary base flow to include periodic base flows for resolvent analysis. This extension enables an improved understanding of how perturbations in a periodic base flow interact with one another and the base flow itself, elucidating cross-frequency modal interactions. Phase-conditioned localized spectral proper orthogonal decomposition (PCL-SPOD) \citep{Franceschini22} and cyclostationary spectral proper orthogonal decomposition \citep{heidt2023spectral} both consider proper orthogonal decomposition within a phase-locked, periodic system. \citet{Franceschini22} leverages a windowed spectral proper orthogonal decomposition technique, using short-time Fourier transforms, to study the spectral proper orthogonal decomposition modes of small-scale turbulent activity within a larger phase variation of a periodic system; e.g. vortex shedding. They also develop a quasi-steady approach that allows this technique to be computed more simply when the large-scale frequencies are much lower than the small-scale frequencies. \citet{heidt2023spectral} also extend spectral proper orthogonal decomposition to globally periodic systems and identify which spectral modes most strongly interact. The approach offers a way of identifying how modes at different frequencies interact with each other and an underlying periodic base state.
%, as a function of the phase of that base state. 
\citet{heidt2023spectral} also demonstrated that cyclostationary POD is the data-driven analog to the harmonic resolvent formulation of \citet{padovan2020analysis}.  These techniques, which focus on decompositions around an intrinsic periodic large scale also share commonalities to the work of \citet{PhasePOD_Fabien}, who utilized phase averaging around a periodic forcing to understand modal response linked to external periodic stimuli.

We aim to build on existing work by introducing a new formulation, intrinsic phase-based POD (IPhaB POD), that enables a near-periodic large scale to be explicitly accounted for in the decomposition. Instead of requiring a known form of the large-scale dynamics, IPhaB POD utilizes a phase portrait from a user-provided low-dimensional proxy for the large-scale behavior. Characteristic periodic behavior is defined from this dynamical systems representation, and cycles from the data that fall within this representative behavior are systematically selected. A POD formulation is constructed for these cycles that are classified as near-periodic, within the time domain. Working within the time domain avoids the need to represent the large scale at a single frequency, since for many flows of interest the large scale is not exactly periodic. The provided time domain formulation uses a clear notion of the phase of the large scale, defined as a fraction of the trajectory along the full cycle in the phase portrait representation. This phase portrait approach allows for cycles with disparate time periods or large-scale trajectories to be meaningfully compared, provided that they correspond to the same near-periodic behavior of interest.  %\textbf{Add references about dynamical systems and trajectories?} \textbf{Add references about periodic perspectives / structures / trajectories in turbulence?}

 As a consequence of the time-domain formulation, the modes generated from IPhaB POD evolve in time rather than in the frequency domain. Utilizing the time domain  could increase representation efficiency and avoid a more broadband representation in frequency space when the underlying large-scale motion (and the modes connected to it) are not perfectly periodic. Nevertheless, this way of defining large-scale phase may be of interest to frequency-based techniques as well, for defining the large-scale position about which the smaller scale modes are defined. We will later  draw connections between IPhaB POD and conditional space-time POD \citep{schmidt2019conditional}, where the event condition here is an evolving large scale that is intrinsic to the system dynamics. We show below that this large scale-based condition requires substantive changes to the POD formulation.

 In this paper we apply IPhaB POD to two examples: a flat plate experiencing vortex shedding and a periodically moving shock front over a cone-cylinder \citep{Duvvuri}. Both examples show that IPhaB POD is able to better capture the large-scale dynamics in the first mode compared with traditional ``space-only'' POD. Moreover, for the cone-cylinder shock example, since the large-scale dynamics are better isolated within the first mode, the next modes better represent small-scale structures that influence and are modulated by the shock undulation process.

\section{Data sets}
\label{sec:data_overview}

To develop and demonstrate the method discussed in this manuscript, two data sets are used. The first is from a high-fidelity computation of an incompressible low Reynolds number flow with only one dominant coherent feature that is periodic (to within discretization errors). The second is from an experiment of a high Reynolds and Mach number flow that has multiple dynamic behaviors interacting with one another. The first is used to introduce the method and its properties, and the second to demonstrate the promise of the method on more complex data. 

\begin{figure}
    \centering
    \includegraphics[width=\textwidth]{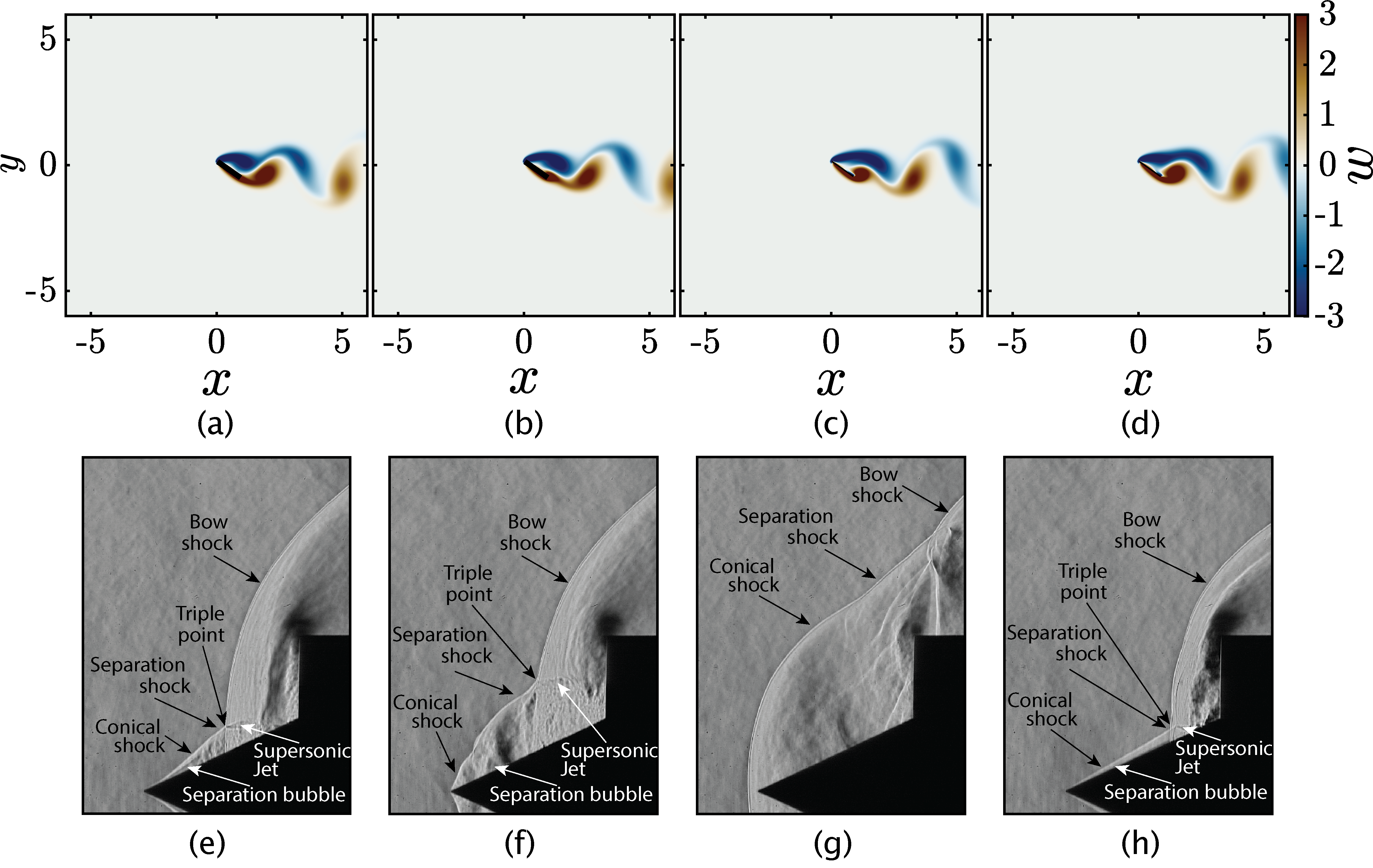}
    \caption{Four instances of (a-d) vortex shedding from flat plate at a stalled angle of attack and (e-h) shock pulsations over a cone-cylinder body \citep{Duvvuri}. Vorticity from a 2D computation is plotted in (a-d). Experimental schlieren data are shown in (e-h).}
    \label{fig:overview_prob}
\end{figure}

The low Reynolds number case is of flow  past a flat plate at an angle of attack of $\alpha = 35^{\circ}$, at a Reynolds number based on the plate length of $Re=100$. At this high (stalled) angle of attack, the long-time behavior after an initial transient is a globally stable limit-cycle attractor in which there is alternating vortex shedding from either side of the body. This periodic limit cycle is shown in figures \ref{fig:overview_prob}(a-d), which show vorticity at four phases in the periodic shedding cycle. The solution for this case was computed using the immersed-boundary method of \citet[]{colonius2008fast}. The simulation parameters are given non-dimensionally as (normalized using the plate length and freestream flow speed) $\Delta x=0.04$ and $\Delta t=0.001$. The simulation was run for a dimensionless time of $t=130$, or 30 vortex-shedding cycles. Over the final ten vortex-shedding cycles, the cycle-to-cycle change in the lift coefficient was less than $1.27 \times 10^{-5}\%$, indicating limit cycle behavior.

The data set involving more complex dynamics, used to demonstrate the promise of the method, comes from the schlieren experiments by \citet{Duvvuri} of high-speed flow past a cone-cylinder body. The dynamics of the flow field are controlled by three non-dimensional geometric parameters: the Mach number ($M$), the ratio $L/D$ of the cone length ($L$) to the diameter of the cylinder base ($D$), and the half angle of the cone ($\theta$). The parameters chosen for this study are $M=6$, $L/D=0.5$, and $\theta=25^\circ$. For the IPhaB POD analysis, we used time-resolved schlieren courtesy of \citet{Duvvuri} with a sampling frequency of 57 kHz. This data was collected for 0.0983 seconds. For these parameters, large-scale pulsations in the shock location are observed. Four instances of the shock pulsations are shown in figures \ref{fig:overview_prob}(e-h), highlighting the various shocks, their interactions, as well as key  processes tethered to that shock interplay (boundary layer development and separation, and the supersonic jet near the triple point).  %We provide a brief summary of the flow physics resulting in the shock pulsations here for context but more information regarding the data and the system's physics can be found in  \citet[]{Duvvuri}. The shock pulsations arise due to complex interactions between a leading bow shock and conical shock arising due to the presence of the base cylinder and conical forebody respectively. The intersection of the conical and bow shocks results in a transmitted shock which impinges on the boundary layer flow on the cone surface. The adverse pressure gradient across this shock results in the formation of a separation bubble which results in a separation shock with a shock angle larger than the conical shock. Thus far, starting from the tip of the cone to the edge of the cylinder, the shock system consists of a conical shock, separation shock, and a bow shock. The intersection point of separation, bow, and transmitted shock is referred to as the triple point and downstream of the transmitted shock, a supersonic jet forms.  The size of the separation bubble due to the adverse pressure gradient continues to increase due to continuous mass accumulation and results in the separation point moving upstream on the cone surface, the separation shock angle growing, and the triple point moving upward. This process continues until the separation shock shape changes and begins to resemble a bow shock. As the conical and separation shock structures change, the bow shock remains the same. As the triple point moves upward, it also rapidly moves downstream and the separation shock transforms into a bow shock. When this transformation is complete, the entire cone-cylinder body is encompassed by the bow shock and this is also associated with a rapid reduction of the separated region. As the separation region decreases in area, the bow shock moves closer to the base cylinder, the conical and separation shocks are reformed, and the cycle restarts. The unsteady quasi-periodic shock pulsations with a identifiable intrinsic large scale, the shock front, and multiple scales of importance present makes this an ideal demonstration problem.
We provide a brief summary of the flow physics resulting in the shock pulsations here for context, but more information regarding the data and the system's physics can be found in  \citet[]{Duvvuri}. The shock pulsations arise due to complex interactions between a bow shock and conical shock. The intersection of the conical and bow shocks results in a transmitted shock which impinges on the boundary layer flow on the cone surface, generating a separation bubble and separation shock with a shock angle larger than the conical shock. The shock front consisting of the conical, separation and bow shock will be denoted as the shock system. Downstream of the transmitted shock, a supersonic jet forms as seen in figures \ref{fig:overview_prob}(e, f and h). The size of the separation bubble increases and results in the separation point moving upstream on the cone surface, the separation shock angle growing, and the triple point moving upward (c.f. figure \ref{fig:overview_prob}(f)). Eventually, the entire cone-cylinder body is encompassed by the shock system (c.f. figure \ref{fig:overview_prob}(g)). The separation bubble then collapses, the shock system rapidly moves closer to the base cylinder, and the cycle restarts (c.f. figure \ref{fig:overview_prob}(h)). This unsteady process is loosely quasi-periodic; i.e., there are shock pulsations with an identifiable intrinsic large-scale that repeats cyclically (the shock motion), but with imperfect periodicity because multiple important scales yield variations in the behavior across each cycle.

 \section{Formulation for intrinsic phase POD}

The proposed formulation aims to decompose a data set that is tethered to and dominated by the phase of some intrinsic large-scale dynamics. We will define a ``large scale'' that controls the overall trajectory of the system, but will allow that large scale and the overall dynamical system to be imperfect in its periodicity. %We define an inner product that aligns different flow realizations with matched phases of the large-scale dynamics of our system. 
The large-scale dynamics will be assumed to have a nominal period duration, $\bar{T}$, with allowed deviations from this nominal trajectory. Because the large-scale is assumed to be near-periodic rather than perfectly periodic, a naive attempt of defining phase as a fixed number of time steps along some $k^{th}$ cycle of the system is inappropriate. Inevitably, the second sample from cycle $k$ will align with a different phase of the large scale from the second sample from cycle $m$ unless care is taken to ensure that they are taken at the same phase of the large-scale dynamics. This challenge was observed by \cite{Franceschini22}: in performing their POD formulation to analyze how a bluff body wake modified turbulent small scales, the authors identified a lack of periodicity in the large scale as an inhibitor to computing meaningful modes.
% that do not retain spurious influences from the imperfect periodicity of the large scale. 
Even for perfectly periodic flows, the need for finite sampling in time can cause a similar outcome: unless the signal is sampled at exactly the same instances along a large-scale period (or with sufficient resolution that the same instances along the cycle can be extracted), the same spurious outcome of having modes inherit (an artificially introduced) non-periodicity of the large scale will occur.

We propose a POD formalism that directly acknowledges the imperfect periodicity of the large scale, and defines an inner product that matches each realization to a common phase of a nominal large-scale, (near-)periodic orbit. The results demonstrate an improved ability to isolate the large-scale behavior from the smaller-scale dynamics that are driven by it. In this section, we characterize the form assumed for the large-scale dynamics, define an expectation operator and inner product informed by this large-scale motion (accounting for its quasi-periodicity), and describe the implementation of the resulting algorithm.

 \subsection{Characterizing the quasi-periodic large-scale dynamics}
 \label{sec:large-scale-deep-dive}
To probe the quasi-periodic large-scale dynamics, we work within a spatially discrete formulation of the governing equations, which we write generically as
\begin{equation}
    \bm{M}\dot{\bm{q}}=\bm{f}(\bm{q}).
    \label{eqn:dyn_sys_gen}
\end{equation}
In (\ref{eqn:dyn_sys_gen}), $\bm{q}$ conveys the flow state sampled at a subset of points, $n_s$,  within the flow domain $\Omega$; e.g., in an incompressible setting $\bm{q}$ is a concatenation of the spatially discrete flow state variables (e.g., for an incompressible flow these would be the various velocity components and pressure at the collection of discrete spatial points used to represent $\Omega$). The vector $\bm{f}$ is the dynamical evolution operator. The matrix $\bm{M}$ allows for non-unity weightings on the time derivatives, including a singular $\bm{M}$ as is needed to represent conservation of mass in the incompressible Navier-Stokes equations (resulting in a differential-algebraic system of Hessenberg index 2). 

We assume the dynamics (\ref{eqn:dyn_sys_gen}) contain underlying large-scale behavior that is quasi-periodic, and that the user has a means of identifying distinct cycles of this behavior. That is, IPhaB POD requires that an operation $\boldsymbol{\mathcal{F}}$ be known such that $\bm{q}_l=\boldsymbol{\mathcal{F}}(\bm{q})$ provides a meaningful way to probe the behavior of the intrinsic large scale. Note that $\bm{q}_l$ does not need to be an approximation of the large scale itself. It only needs to undergo dynamics that allow for the distinct cycles of the large scale, and the large-scale phase along each cycle, to be identified. We refer to the dynamics of $\bm{q}_l$ (i.e., the dynamics of $\bm{q}$ under $\boldsymbol{\mathcal{F}}$)  as restricted dynamics.  While the requirement of an appropriate $\boldsymbol{\mathcal{F}}$ may appear somewhat limiting, in many cases such knowledge of the large scale is available. Examples of flows with intrinsic large-scale dynamics include wakes, separation bubble ``breathing,'' fluid-structure interaction problems, and rotating systems. We will mathematically characterize the assumed large-scale behavior and then give examples of this mathematical representation for the two representative cases described in section \ref{sec:data_overview}.

We assume the restricted dynamics, $\bm{q}_l=\boldsymbol{\mathcal{F}}(\bm{q})$, yield $n_c$ cycles that can be used for this analysis. We further assume that the restricted trajectory associated with the $k^{th}$ cycle, $\{\bm{q}_l(t): t\in [t_{0_k}, t_{f_k} ]\}$, $k \in [1, 2, \dots, n_c]$, has a (near-)period $T_k:= t_{f_k}- t_{0_k} $. Associated with each $k^{th}$ cycle is also a minimum and maximum at say, $t_{min_k}$ and $t_{max_k}$, given by derivative conditions of $\bm{q}_l$ over that cycle. The peak-to-peak amplitude for each cycle can thus be defined as $A_k := ||\bm{q}_l(t_{max_k})-\bm{q}_l(t_{min_k})||$. One can define a nominal period duration and peak-to-peak amplitude from the arithmetic mean of the quantities $T_k$ and $A_k$ over all cycles; e.g., $\bar{A}= 1/n_{c}\sum_{k=1}^{n_c} A_k$ (with an analogous definition for $\bar{T}$).

Because the dynamics are assumed to be imperfectly periodic, in general $\bm{q}_l(t_{f_k})\ne \bm{q}_l(t_{0_k})$. In fact, there may be cycles for which the period duration is very different from the nominal period $\bar{T}$, and the restricted trajectory $\{\bm{q}_l(t): t\in [t_{0_k}, t_{f_k} ]\}$ yields large values of $||\bm{q}_l(t_{f_k})- \bm{q}_l(t_{0_k})||$. For the purposes of this article, we focus on restricted trajectories that exhibit near-periodic orbits, and omit extreme events from consideration. That is, we only perform IPhaB POD on cycles that are near-periodic in the sense that their restricted trajectories in phase space are closed and their associated time durations are equal to a representative period duration (to within some tolerance). The near-periodic trajectories considered within IPhaB POD are then computed as those satisfying
\begin{equation}
%\begin{split}
\gamma_{j} = \{\bm{q}_l(t): t \in [t_{0_{j}},t_{f_{j}}], |\bar{T}-T_j|/\bar{T} := \delta_j<\delta^*,||\bm{q}_{l}(t_{f_{j}}) - \bm{q}_{l}(t_{0_{j}})||/\bar{A} := \epsilon_j <\epsilon^*\},
\label{eqn:near-period-def}
%\end{split}
\end{equation}
where $\delta^*$ and $\epsilon^*$ are threshold values for departure from the nominal period duration and from trajectory closedness. We denote the number of near-periodic orbits satisfying (\ref{eqn:near-period-def}) as $n_p$, the set of indices associated with these orbits as $N$ (with $dim(N)=n_p\le n_c$), and $\Gamma=\{\gamma_j\}, j\in N$.

It will be useful in presenting IPhaB POD to define the ``most'' representative periodic cycle of the restricted dynamics, which we denote as the trajectory $\gamma^p :=  \{\bm{q}_l(t): t\in [t_{0_k}, t_{f_k} ]\}$, $k\in N$, that satisfies $\min_{k} \sqrt{\delta_k^2 + \epsilon_k^2}$. (Note that, by construction, $\gamma^p$ is one of the admitted near-periodic trajectories in (\ref{eqn:near-period-def}).)  We denote the period and discrepancy from trajectory closedness of the nominally periodic $\gamma^p$ as $T^p$ and $\epsilon^p$, respectively.

We now discuss how the trajectories $\gamma_j$, $j\in N$, and $\gamma^p$ are obtained for the two problems considered in this article. In addition to illustrating how these trajectories are determined, the examples are also intended to motivate as modestly reasonable the assumption that a proxy for probing large-scale cycles, given by a user-provided mapping $\boldsymbol{\mathcal{F}}$, be known. 

For the aerodynamic flat plate problem, % , the long-time behavior after an initial transient is a globally stable limit-cycle attractor in which there is alternating vortex shedding from either side of the body. This periodic limit cycle is shown in figure \ref{fig:overview_prob}(a-d), which shows vorticity at four phases in the periodic shedding cycle.
vortex shedding is the large (only) scale (c.f., figures \ref{fig:overview_prob} (a-d) and the surrounding discussion) and the lift dynamics can be used to characterize the large-scale vortex-shedding behavior. That is, while the lift is not equivalent to the large-scale vortex shedding, its dynamics can be used to identify distinct phases of this vortex-shedding behavior. Thus, for this problem, an $\boldsymbol{\mathcal{F}}$ can be defined that takes the flow state, extracts the flow stresses via the gradient of the pressure field and the viscous Laplacian of the flow velocity, and evaluates these on the surface of the aerodynamic body. (Or alternatively, many immersed boundary methods provide the surface stresses directly as a state variable, and under this formulation the surface stress state variable may be spatially integrated.) 

\begin{figure}
    \centering
    \includegraphics[width=\textwidth]{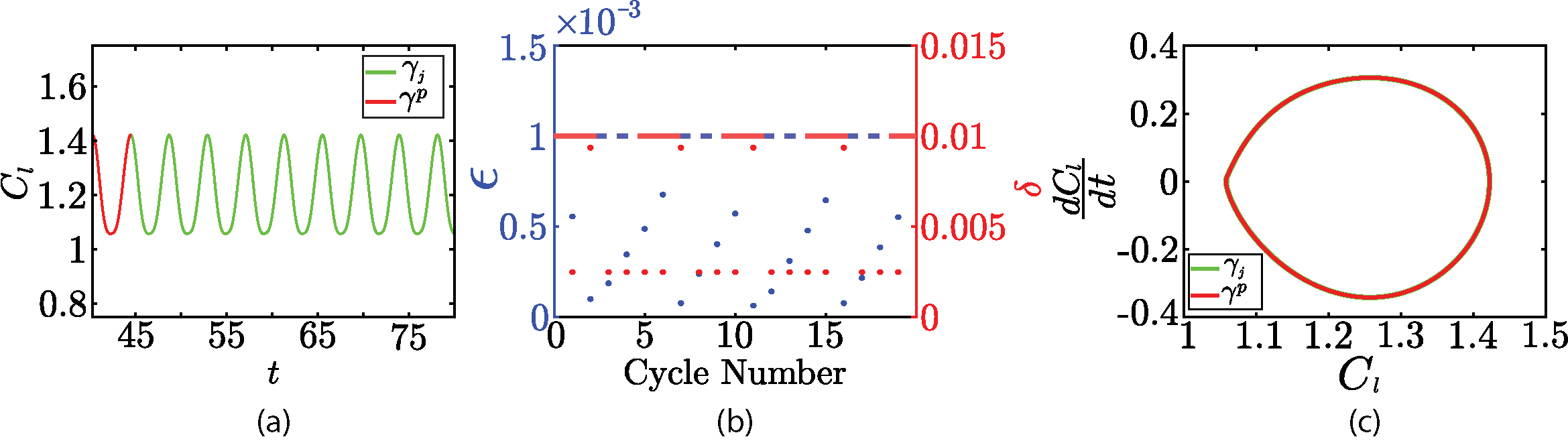}
    \caption{(a) The coefficient of lift, $C_L$, for the representative cycle of vortex shedding from the flat plate is shown in red and the remaining ensembles are plotted in green. (b) Nearness of cycle start and end positions ($\epsilon$) and difference of cycle period to nominal period ($\delta$) for all cycles. Threshold values used to determine near-periodic cycles shown in dashed lines (for this case, all cycles are nearly periodic). (c) Cycles plotted using a phase portrait with the coefficient of lift plotted against its time derivative. All cycles lie on top of each other because all are periodic. }
    \label{fig:FP_LS_identify}
\end{figure}

The periodic trajectories $\Gamma$ can be extracted from this definition of $\boldsymbol{\mathcal{F}}$ as follows (and as illustrated in figure \ref{fig:FP_LS_identify}). The lift signal depicted in figure \ref{fig:FP_LS_identify}(a), plotted sufficiently far after the initial transient that the limit cycle has been reached, demonstrates the periodic behavior of the system. The variation in trajectory closedness of the lift cycle and discrepancy from the nominal period duration is shown in figure \ref{fig:FP_LS_identify}(b). For this problem, the cycle-to-cycle differences are small and due to discretization errors, so thresholds of $\epsilon^* = 0.001$ and $\delta^* = 0.01$ are chosen to include all cycles. For illustration purposes, and to indicate a phase trajectory undergone by the lift as a  representation of the large scale, figure \ref{fig:FP_LS_identify}(c) shows the various lift cycles superimposed on one another as a phase portrait of $dC_l/dt$ versus $C_l$, though only a single curve is visible because of how nearly periodic the system is. The cycle depicted in red in figures \ref{fig:FP_LS_identify}(a,c) is the one computed as $\gamma^p$ based on the criterion described above, though again the periodicity of the system dynamics render the specific choice of $\gamma^p$ non-essential for this problem. Each distinct near-periodic trajectory within $\Gamma$, defined through the above procedure, indicates a distinct vortex-shedding cycle. Moreover, the trajectory exhibited by each $\gamma_j\in\Gamma$ experiences each of the different phases of the vortex-shedding process within a single cycle.  

Note that at higher Reynolds numbers, the vortex shedding of flow past a flat plate would lose perfect periodicity due to turbulent fluctuations, though it would retain a strong signature at the characteristic vortex-shedding frequency which is robust to Reynolds number when scaled appropriately \citep{roshko1954drag}. For this higher Reynolds number case, one may wish to modify the $\boldsymbol{\mathcal{F}}$ operation to not only compute the lift but also remove the high-frequency components associated with the turbulent fluctuations. This low-frequency signal could be used to establish the large-scale vortex-shedding behavior that distinguishes the various cycles. 

\begin{figure}
\centering
\includegraphics[width=\linewidth]{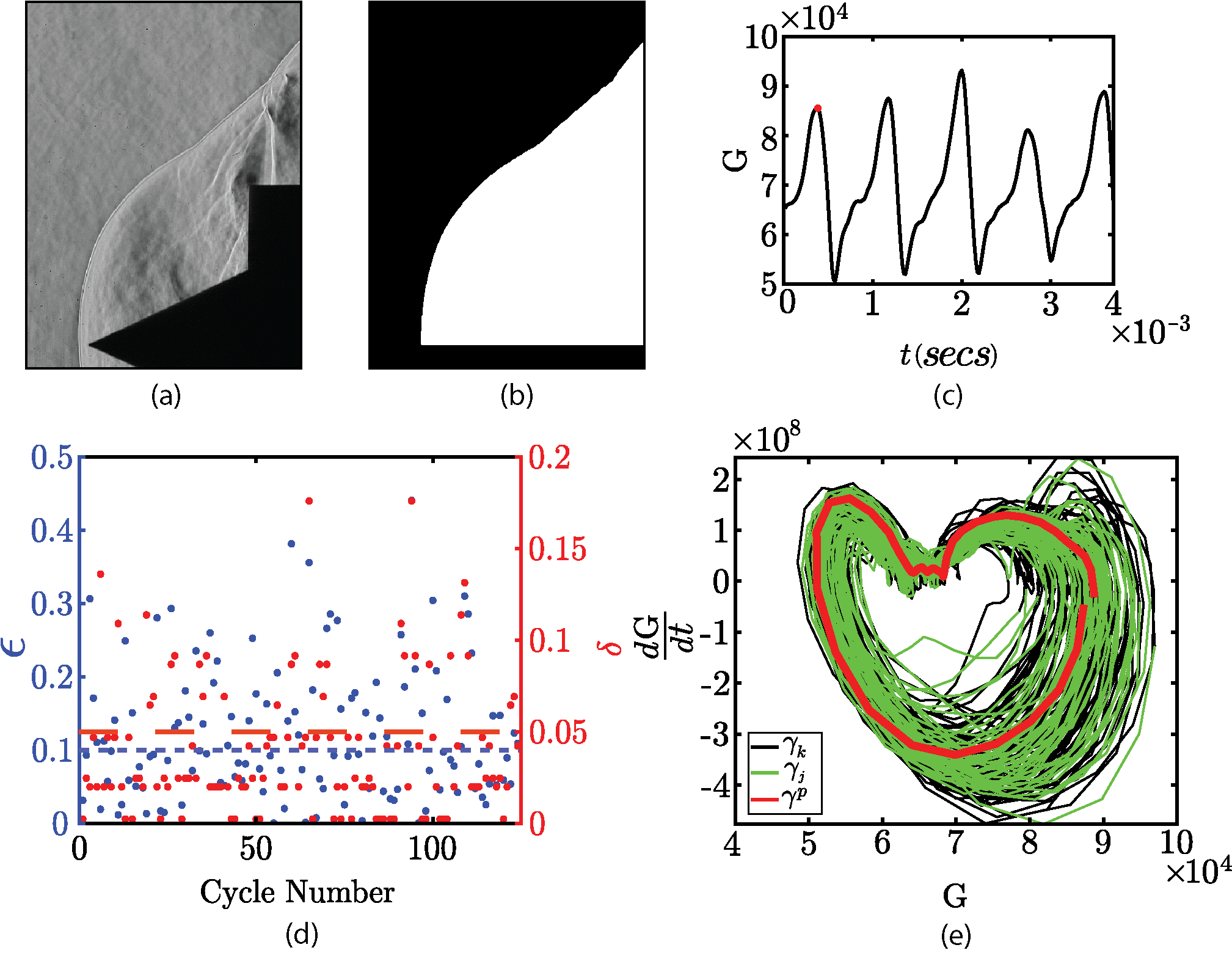}
     \caption{(a) Schlieren image of shock oscillation over cone-cylinder body. (b) Filtered version of (a) to identify phase of large scale. (c) Spatially integrated value of the filtered image shown in (b) plotted against time, with the time instance of (a) marked in red. (d) Nearness of cycle start and end positions ($\epsilon$) and difference of cycle period to nominal period ($\delta$) for all cycles. Threshold values used to determine near-periodic cycles shown in dashed lines. (e) Cycles plotted using a phase portrait with the value of plot (c) plotted against its time derivative. Cycles that are considered near-periodic using the thresholds of (d) are shown in green, while the representative cycle is shown in red. }
    \label{fig:shock_LS_identify}
\end{figure}

For the shock problem, the large scale is the shock, which has an undulating location (c.f., figures \ref{fig:overview_prob}(e-h)). For this problem, our choice of $\boldsymbol{\mathcal{F}}$ leverages an image filter: an edge-detection algorithm is used to define the upstream (downstream) region of the shock, labeling this region with a value of zero (unity). An unaltered schlieren image is shown in figure \ref{fig:shock_LS_identify}(a) with the filtered image using the edge-detection algorithm for that same image shown in figure \ref{fig:shock_LS_identify}(b). The filtered image is then integrated in space to yield a one-dimensional signal that varies in time, shown in figure \ref{fig:shock_LS_identify}(c). (The time instance shown in figures \ref{fig:shock_LS_identify}(a,b) is indicated by the red marker in figure \ref{fig:shock_LS_identify}(c).) This spatially integrated signal is an indicator of the large-scale dynamics, as its value changes as the shock moves within the domain.  
 Similar to the flat plate problem, to indicate phase trajectories undergone by the integrated signal from figure \ref{fig:shock_LS_identify}(c), figure \ref{fig:shock_LS_identify}(e) depicts the integrated signal against its time derivative. 
 
 Visible in both figures \ref{fig:shock_LS_identify}(c,e), is an apparent lack of periodicity that nonetheless contains common cyclic behavior (with significant cycle-to-cycle variation). This cycle-to-cycle variation is quantitatively demonstrated in figure \ref{fig:shock_LS_identify}(d) through the difference between each cycle's start and end value ($\epsilon$) and the difference between each cycle's period and a reference period ($\delta$). For this problem, the thresholds of these values that are used to define near-periodic orbits were chosen to be $\epsilon^* = 0.1$ and $\delta^*= 0.05$.
%This choice of threshold values is somewhat arbitrary; future work includes continued refinement of how these parameters are chosen. 
These thresholds were chosen as a balance between (1) removing extreme events with significant differences from the nominal trajectory based on closedness and period length and (2) over trimming cycles with some differences due to cycle-to-cycle variations that are important to understand. The characteristic periodic cycle, $\gamma^p$, is highlighted in red in \ref{fig:shock_LS_identify}(e), while all cycles are shown in black and all cycles that are the near-periodic cycles belonging to $\Gamma$ are shown in green. The computed trajectories $\Gamma$ contain information about the shock motion for cycles where the undulations remain near a representative periodic orbit. As such, each restricted trajectory $\gamma_j\in\Gamma$ represents a distinct cycle of the shock's oscillation process, and the trajectory exhibited by each of these cycles reflects the various shock locations explored through an oscillation cycle.%As demonstrated in figure \ref{fig:shock_LS_identify},  the resulting integrated signal exhibits quasi-periodic dynamics reflective of the near-periodic motion of the shock. %A few practical implementation points: we use a Gaussian function to smooth the edges defined by the edge-detection algorithm, given by
% \begin{equation}
%     \text{define Gaussian}
% \end{equation} 

Note that the data we utilize for the shock problem comes from schlieren imaging, so that the filtering operation defined above is not performed on the flow state $\bm{q}$ directly but on integrated observations of refractive index changes induced by density gradients in the flow. We do not focus here on this distinction; one could envision an analogous $\boldsymbol{\mathcal{F}}$ operation that acts on the full compressible flow state, identifying discontinuities in any of the flow quantities and using this location to label upstream and downstream regions and integrate the result over the spatial domain as indicated above.

We highlight that while in both examples the operation $\boldsymbol{\mathcal{F}}$ is chosen to map from the flow state to a real scalar value, this choice is not unique (nor is the choice of mapping, even when restricting the image of the operation to the collection of real numbers). Indeed, mappings onto higher dimensional outputs might facilitate a more robust analysis of the large-scale behavior described below. As the main focus of this article is on utilizing a relevant representation of the intrinsic large-scale on which to perform POD,  we leave the topic of optimal large-scale extraction as a point of future research.

\subsection{A large scale-informed POD} \label{sec:SPPOD-math-formulation}

This section describes how to use the collection of near-periodic trajectories $\Gamma$, associated with the restricted dynamics, to define an expectation operator and inner product that compare distinct cycles along matched phases of the large scale.
The result is an energy-ranked modal decomposition, with all the inherited optimality properties of POD, that are informed by this large-scale behavior.

The expectation operator is defined as an ensemble average over the collection of full flow state trajectories which, under the map $\boldsymbol{\mathcal{F}}$, belong to the near-periodic trajectories $\Gamma$. That is, the sample space being drawn from is $j\in N$ and we may, for clarity, explicitly write the full flow state as a random variable or a function of this sample space via
\begin{equation}
    \tilde{\bm{q}}(t, j) = \bm{q}(t), \text{ for } t\in[t_{0_j}, t_{f_j}], j\in N, %\{\bm{q} : \mathcal{F}(\bm{q}) \in \gamma_j \}
\end{equation}
where the dependence on the sample space is explicitly stated. 

Intuitively, we seek an expectation operator defined roughly as an ensemble average over these realizations
$
    \langle \tilde{\bm{q}} \rangle(t) = \frac{1}{n_p} \sum_{j\in N} \tilde{\bm{q}}(t, j)
$. (Recall $n_p=dim(N)$). One issue with this form is that it is not possible to evaluate this expectation operation at a given time instance, since each $\tilde{\bm{q}}(\cdot, j)$ is only defined over the time interval $t\in[t_{0_j}, t_{f_j}]$. More substantively, even barring that challenge, comparing across the same time within a cycle does not generally correspond to the same phase of the large-scale motion, because of the deviations from periodicity of the large scale. 

To address both of these issues, we define a prototypical large-scale motion and relate the time interval of each of the cases with near-periodic cycles $j\in N$ to the nearest time instance of the reference prototypical cycle. This characteristic trajectory is chosen as $\gamma^p$, defined in section \ref{sec:large-scale-deep-dive},  for which the associated $\sqrt{\delta^2+\epsilon^2}$ is minimized (i.e., minimizing the difference from the nominal period length and the departure from trajectory closedness).   

To enable a sensible averaging operation expressed in terms of a unified time interval, we require a map from each time instance $t\in[t_{0_j}, t_{f_j}]$ to the time instance along the nominal trajectory, $t^p\in[t_{0}^p, t_{f}^p]$, for which the large-scale dynamics are most similar. Denoting $\tau_j = [t_{0_j}, t_{f_j}]$ and $\tau^p = [t_{0}^p, t_{f}^p]$, we define this map $\mathcal{T}_j^p: \tau_{j} \rightarrow \tau^p$ using
\begin{equation}
t^p = \min_{t' \in [t^p_0, t_f^p]} \norm[\bm{q}_l^p(t') - \bm{q}_{l_{j}}(t)], \; t\in\tau_j.
\label{eqn:t_map}
\end{equation}
There is a related map $\mathcal{T}_j^p: \tau^{p} \rightarrow \tau_j$, where $\mathcal{T}_j^p = (\mathcal{T}_p^j)^{-1}$ provided that $\mathcal{T}_p^j$ is injective. Intuitively, this requirement is modest: the trajectories are selected to represent nearly the same phenomenon, so supposing a one-to-one mapping from one time interval to the other is natural. 

Using (\ref{eqn:t_map}), we arrive at an expression for the expectation operator, given by
\begin{equation}
     \langle \tilde{\bm{q}} \rangle(t) = \frac{1}{n_p} \sum_{j\in N} \tilde{\bm{q}}(\mathcal{T}_j^p(t), j), \quad t \in \tau^p.
     \label{eqn:expectation}
\end{equation}
Note that this definition of an expectation operator is strongly connected to the conditional expectation defined in \citet{schmidt2019conditional}. Indeed, the expression (\ref{eqn:expectation}) entails a conditional expectation operator. Unlike in \citet{schmidt2019conditional}, however, the expectation here is not conditioned on a specific event, but rather on a phase of an intrinsic large scale which evolves over time in a quasi-periodic fashion. This difference has required new formulations to the definition of the realizations developed in (\ref{eqn:near-period-def}) as well as a time mapping that facilitates a clear comparison of the dynamics of each realization relative to an underlying large-scale behavior. This time mapping directly affects the definition of the expectation operator in (\ref{eqn:expectation}) and will also manifest itself in the definition of the inner product (described next).

With the expectation operator defined, POD can be performed once a suitable inner product is chosen. Similar to the expectation operator, we emphasize that an inner product that simply integrates along the trajectory of each realization, $\tilde{\bm{q}}(\cdot, j)$ ($j\in N$), would not account for the imperfect periodicity in the large scale associated with that trajectory, and would yield a modal representation which erroneously inherits features of the large scale in ``higher modes'' that are tethered to this large-scale motion. We therefore choose to perform this trajectory integration relative to a nominal large-scale trajectory. Each instance of the trajectory encoded in $\tilde{\bm{q}}(\cdot, j)$ ($j\in N$) will be cast in terms of the nearest instance on the nominal trajectory. This approach facilitates a meaningful comparison across the various ensembles: all trajectory integrations are aligned with a shared (near-) orbit reflective of the periodic dynamics that each of the ensembles are close to as defined by (\ref{eqn:near-period-def}). Denoting the manifold on which the dynamics $\bm{q}$ evolve as $\boldsymbol{\mathcal{M}}$, we define the inner product as 
\begin{equation}
    \IP[\bm{r}][\bm{m}] = \int_{t_0^p}^{t_{f}^p} \bm{m}^*(\mathcal{T}_j^p(t')) \bm{W} \bm{r}(\mathcal{T}_j^p(t')) dt', \quad \forall \bm{r}, \bm{m}\in \boldsymbol{\mathcal{M}},
    \label{eqn:IP_iphase}
\end{equation}
where $(\cdot)^*$ denotes complex conjugation and $\bm{W}$ is an appropriately chosen positive-definite weighting matrix that incorporates quadrature weights to approximate spatial integration of the space-discrete state variables, and allows for the induced norm to be a meaningful representation of physical energy. (Because of the space-time formulation employed here,  most applications of this method would likely involve real-valued functions. We leave the complex conjugate in the inner product definition for generality). The exact construction of $\bm{W}$ varies from problem to problem, but is generally dependent on the field value chosen. The state variable used for the flat plate problem is vorticity, so the weighting chosen is a diagonal matrix with $W_{jj}=\Delta x_j\Delta y_j$, the space between the $j^{th}$ spatial point and its adjacent one. That is, the resulting norm represents the enstrophy (per unit of density) of the system. The ``state variable'' used in the shock problem is the pixel intensity provided by each schieleren image. The image intensity is driven by density gradients in the flow, and the same weight of $W_{jj}=\Delta x_j\Delta y_j$ is chosen. (Note that the third spatial dimension is neglected for both the 2D flat plate problem and for the 3D shock problem, since for the latter only plane-wise data is available). This choice avoids weighting certain regions of space more than others. While the resulting norm has a non-obvious direct physical interpretation because the pixel intensity is not a flow state variable, it is proportional to the square density gradient field. 

The eigenvalue problem that yields the intrinsic phase-based (IPhaB) POD modes then takes the standard form, obtained by searching for the mode $\bm{\phi}$ that maximizes the expectation of the projection of the flow state onto that mode. The result is 
\begin{equation}
    (\mathcal{R}\bm{\phi})(t) = \langle (\bm{\phi}, \tilde{\bm{q}}) \tilde{\bm{q}} \rangle (t) = \int_{t_0^p}^{t_f^p} \bm{R}(t, t') \bm{W} \bm{\phi}(t') dt' = \lambda \bm{\phi}(t), \quad t \in \tau^p,
    \label{eqn:eval_prob}
\end{equation}
where  $\bm{R}(t, t') = \avg[ \tilde{\bm{q}}\tilde{\bm{q}}^*](t, t')$ is the autocorrelation function. 

Because the proposed formulation utilizes the standard POD formulation for a specific choice of inner product and expectation operator, prior results of operator self-adjointness, optimality, and mode coherence apply. The operator $\mathcal{R}$ is self-adjoint and the eigenvalue problem (\ref{eqn:eval_prob}) yields a collection of modes $\{\bm{\phi}_1, \bm{\phi}_2, \dots\}$ that are orthogonal with respect to the inner product (\ref{eqn:IP_iphase}). The associated set of eigenvalues are positive and real-valued, and are taken here to be ordered such that $\lambda_1 \ge \lambda_2 \ge \cdots$ (see, for example, \citet{Holmes_Lumley_Berkooz_Rowley_2012} for a review of these results). An equation for, say, the $n+1$ most energetic mode can be obtained by maximizing the expectation of the projection of the flow state onto a subspace orthogonal to the span of the first $n$ modes. Optimality of low rank representation and space-time coherence of the modes also hold. For a discussion of the latter point about coherence and the breakdown of this property when inner products that do not incorporate time are used, see \citet{TowneSPOD}. Note that in our case coherence is with respect to time scaled by the nominal large-scale trajectory.

 \subsection{Computing IPhaB POD modes with sampled data} \label{sec:SPPOD-algorithm}

To approximate the time-continuous formulation above with discrete samples of data, we construct a data matrix where the $j^{th}$ column contains the flow state associated with the quasi-periodic trajectory $\gamma_j$, $j\in N$. We recall that, to facilitate a meaningful comparison across the other trajectories,  the eigenvalue problem (\ref{eqn:eval_prob}) is cast in time according to the nominal periodic trajectory $\gamma^p$ rather than the $j^{th}$ trajectory. Denoting by $n_n$ the number of time samples of this nominal periodic trajectory, and recalling that $n_p=dim(N)$ is the number of near-periodic cycles contained within the set $\Gamma$,
the data matrix to be formed is then
\begin{equation}
    \boldsymbol{X}=
    \begin{bmatrix}
    \tilde{\boldsymbol{q}}(\mathcal{T}_{n_1}^p(t_1), n_1) & \tilde{\boldsymbol{q}}(\mathcal{T}_{n_2}^p(t_1), n_2) & \dots & \tilde{\boldsymbol{q}}(\mathcal{T}_{n_p}^p(t_1), n_p) \\
    \tilde{\boldsymbol{q}}(\mathcal{T}_{n_1}^p(t_2), n_1) & \tilde{\boldsymbol{q}}(\mathcal{T}_{n_2}^p(t_2), n_2) & \dots & \tilde{\boldsymbol{q}}(\mathcal{T}_{n_p}^p(t_2), n_p) \\
    \vdots & \vdots & \ddots & \vdots \\ 
    \tilde{\boldsymbol{q}}(\mathcal{T}_{n_1}^p(t_{n_n}), n_1) & \tilde{\boldsymbol{q}}(\mathcal{T}_{n_2}^p(t_{n_n}), n_2) & \dots & \tilde{\boldsymbol{q}}(\mathcal{T}_{n_p}^p(t_{n_n}), n_p)
    \end{bmatrix} .
\label{eqn:XSPPOD}
\end{equation}

Using this definition (\ref{eqn:XSPPOD}) of the data matrix, the eigenvalue problem that approximates (\ref{eqn:eval_prob}) is
\begin{equation}
\bm{X}\bm{X}^*\tilde{\bm{W}}\bm{\phi}_d = \bm{\Lambda} \bm{\phi}_d,
\label{eqn:eval_disc}
\end{equation}
where $\tilde{\bm{W}}$ is a weighting matrix that incorporates the matrix $\bm{W}$ from (\ref{eqn:eval_prob}) as well as quadrature weights to approximate the time integral in the inner product definition.  This weighting matrix is block diagonal, and constructed as a concatenation of $\frac{1}{n_p}\bm{W}$, repeated along block diagonal entries $n_p$ times. This scaling accounts for the ensemble average (and the spatial integral via $\bm{W}$). (For reference, results shown in later sections for the commonly used space-only POD, which forms the data matrix to have each snapshot of the full data set in a separate column, the weighting matrix is $\tilde{\bm{W}} = \frac{1}{n_t}\bm{W}$, where $n_t$ is the number of snapshots used for the space-only decomposition. This number need not equal $n_n\cdot n_p$, since generally space-only POD does not omit snapshots from different cycles, and instead uses the full data set.) The eigenvector $\bm{\phi}_d$ discretely approximates the time-continuous eigenvector $\bm{\phi}$. With the state vector $\bm{q}\in \mathbb{R}^{n_s}$, where $n_s$ is the number of points used to represent the spatial domain $\Omega$, $\bm{\phi}_d\in\mathbb{R}^{n_s\cdot n_n}$. The $k^{th}$ collection ($k=1, \dots, n_n$) of $n_s$ entries in $\bm{\phi}_d$ approximates $\bm{\phi}(t_k)$. In practice the eigenvalue problem (\ref{eqn:eval_disc}) is typically computationally unwieldy, and the method of snapshots \cite[]{Sirovich} may be used to extract the desired modes from the smaller eigenvalue problem associated with $\bm{X}^*\tilde{\bm{W}}\bm{X}$. 

We now detail the process used to compute $\mathcal{T}_{j}^p(t_k)$, where $j=n_1, \dots, n_p$; $k=1, \dots, n_n$. First, we define from the data the nominal large-scale trajectory $\gamma^p=\{\tilde{\bm{q}}(t_{0_l}, l), \tilde{\bm{q}}(t_{1_l}, l) \dots, \tilde{\bm{q}}(t_{f_l}, l)\}$ for the index $l\in N$ that yields the minimal value of $\sqrt{\epsilon^2+\delta^2}$. (Note that this trajectory $\gamma^p$ is the time-sampled variant of the trajectory defined in section \ref{sec:SPPOD-math-formulation}; we avoid using a distinct variable for notational ease.) That is, the time instances indicated in the data matrix (\ref{eqn:XSPPOD}) are defined as $\{t_0, t_1, \dots, t_{n_n}\}:=\{t_{0_l}, t_{1_l}, \dots, t_{f_l}\}$. We then build the data matrix where each column contains a different realization of $j\in N, \; j\ne l$. 

For each $j^{th}$ column, we populate the collection of rows $\tilde{\boldsymbol{q}}(\mathcal{T}_{j}^p(t_k), j)$ by interpolation across the collection of available snapshots over the $j^{th}$ cycle as follows. First, we identify for the $j^{th}$ cycle the time $t_q$ which yields a flow state nearest to the flow state associated with the nominal periodic cycle:
\begin{equation}
    t_q = \argmax_{t'\in\tau^j} \frac{\tilde{\bm{q}}(t_{k}, l)^*\bm{W} \tilde{\bm{q}}(t', j)}{||\tilde{\bm{q}}(t_{k}, l)||^2}.
\end{equation}
Because of finite sampling in time, this nearest flow state in the $j^{th}$ cycle to that associated with the desired phase of the large scale will not in general be desirably close to $\tilde{\bm{q}}(t_{k}, l)$. As such, standard quadratic interpolation is used over the time instances $[t_{q-1}, t_q, t_{q+1}]$ to find the flow state within this time interval that yields the maximal projection onto $\tilde{\bm{q}}(t_{k}, l)$. This interpolated flow state is used to populate $\tilde{\boldsymbol{q}}(\mathcal{T}_{j}^p(t_k), j)$ in (\ref{eqn:XSPPOD}), for $j=n_1, \dots, n_p$; $k=1, \dots, n_n$.

This proposed approach yields, at each cycle (column in the data matrix $\bm{X}$), snapshots that are matched according to the phase of an underlying large scale driving the dynamics. The utility of this approach for cases where the underlying large scale is imperfectly periodic is that it creates a meaningful comparison across cycles with differences that obscure an important underlying similarity (a shared large-scale phase). The proposed approach also has utility in cases where the large-scale dynamics are perfectly periodic. Even in this setting, finite sampling in time will generally yield a collection of snapshots within a cycle that are at different instances along the phase of the large scale, which if compared directly do not enable a meaningful assessment of how the dynamics evolve relative to the motion of the driving large-scale motion.   

\section{Results}\label{Demos}
\subsection{Vortex shedding from a flat plate with $Re=100$, $\alpha=35 ^{\circ}$}
We first demonstrate IPhaB POD using the $Re = 100$ flat plate problem described in section \ref{sec:data_overview}. We aim to illustrate the ability of IPhaB POD to fully capture the system's dynamics in one IPhaB POD mode when there is only one periodic and laminar flow feature in the data set. The same data set used to produce figure \ref{fig:overview_prob}(a-d), and described in section \ref{sec:data_overview}, is used for the analysis of this problem. 

%We used an immersed-boundary method based solver \cite[]{GozaandColonius} to acquire the data of vortex shedding over a flat plate at an angle of attack of $35^{\circ}$ and $Re=100$. The flat plate at a stall angle of attack lead to periodic shedding leading and trailing edge vortices as seen in the instantaneous snapshots in figure \ref{fig:overview_prob}. We used a spatial grid and time-step sizes of $\Delta x=0.04$ and $\Delta t=0.001$ and ran the simulation for an adequate time to ensure the limit-cycle dynamics were reached. Figure \ref{fig:overview_prob} contains four instantaneous snapshots of the vorticity field where the vortex shedding from the leading and trailing edges of the flat plate is visible. The leading and trailing edge vortices were the only scales present in this flow field and therefore we identified them to be the the large scale for this problem. 
As discussed in section \ref{sec:large-scale-deep-dive}, the near-periodic trajectories for the analysis $\Gamma$ were computed using information about the plate coefficient of lift, $C_l$. Cycles for the lift dynamics were obtained using the local maxima in $C_l$. The nominal period length of each cycle was computed to be 4.211, and 84 distinct phases were computed along each cycle; i.e., $n_n= 84$. Nineteen cycles were collected after the initial transient, and due to the system periodicity all cycles were kept for the IPhaB POD analysis (see section \ref{sec:large-scale-deep-dive}). In the ensuing discussion, IPhaB POD results are compared to those for space-only POD (for the latter, to provide a straightforward comparison we did not subtract the temporal mean). 

\begin{figure}
    \centering
    \includegraphics[width=\textwidth]{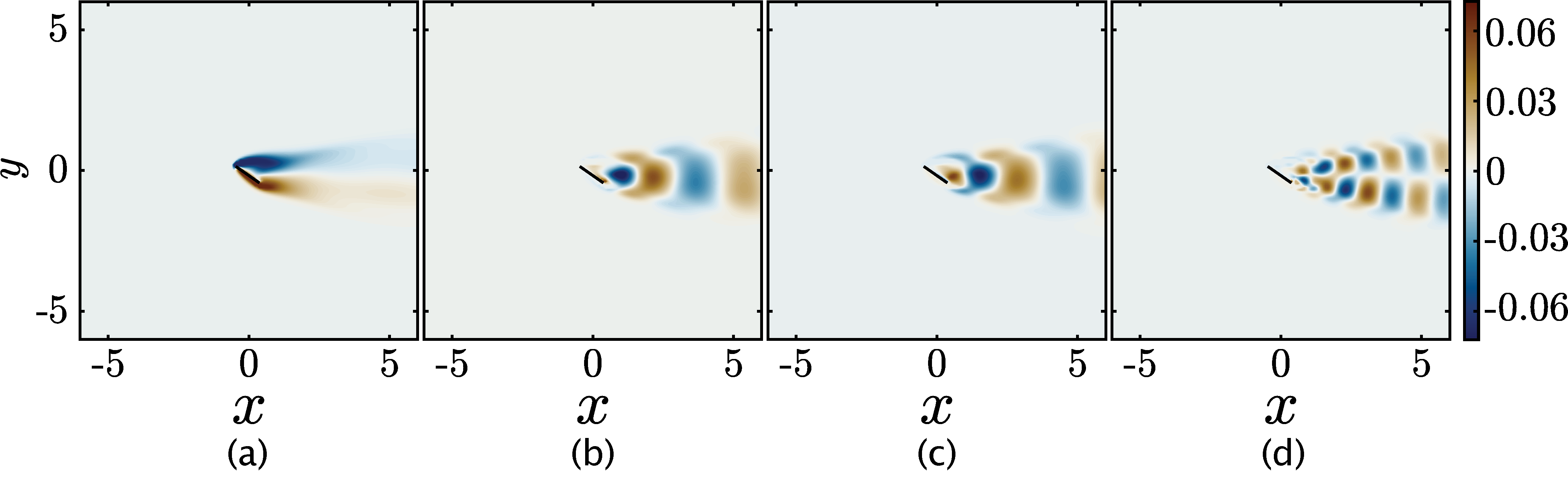}
    \caption{(a) First, (b) second, (c) third, and (d) fourth space-only POD modes for vortex shedding over a flat plate at stall angle of attack.}
    \label{fig:FP_POD}
\end{figure}

\begin{figure}
    \centering
    \includegraphics[width=\textwidth]{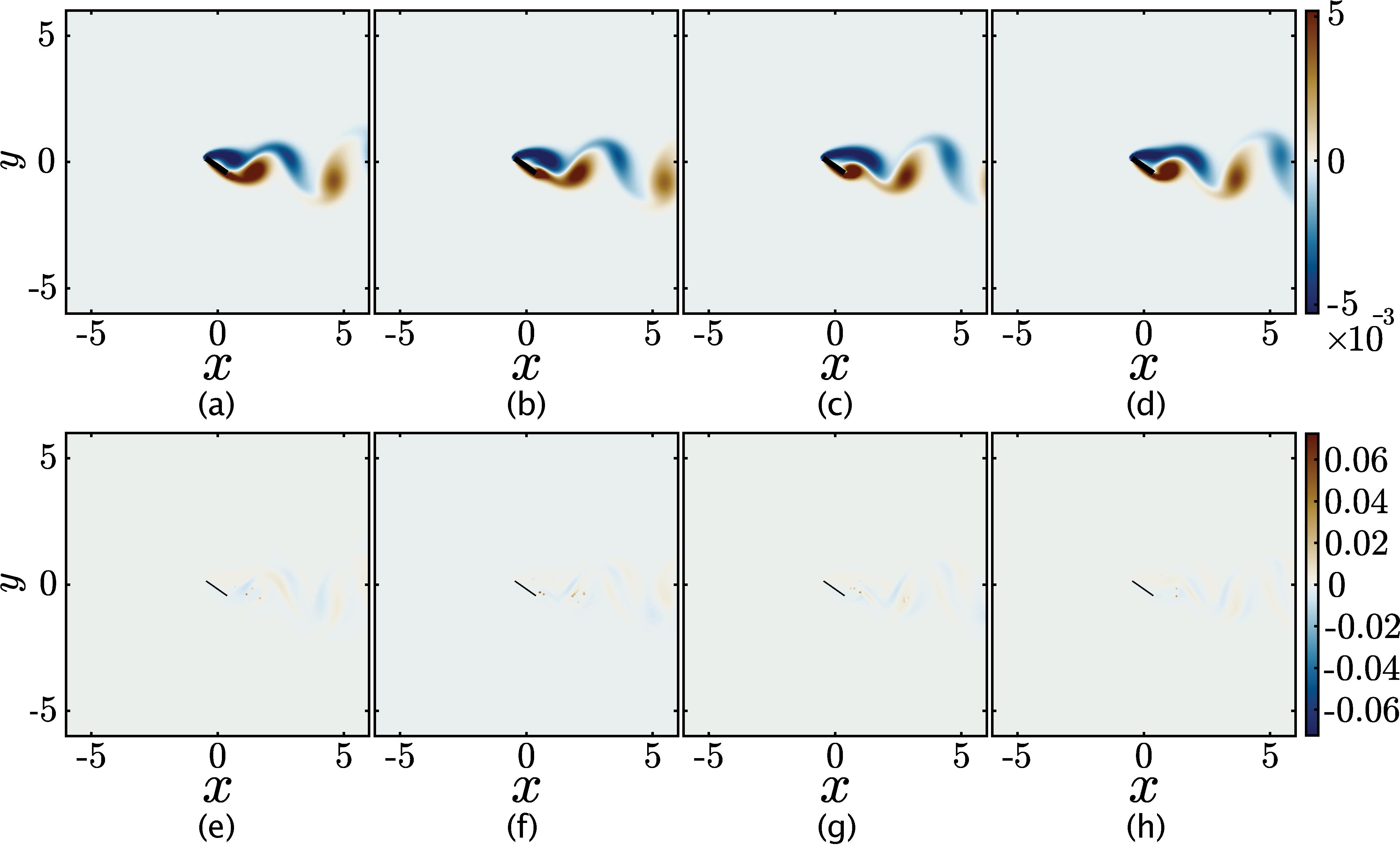}
    \caption{Four equally distributed phase instances (1st (a,e), 21st (b,f), 42nd (c,g), and 63rd(d,h) out of a total of 84 distinct phases) of the (a-d) first, and (e-h) second IPHaB POD modes for vortex shedding over a flat plate at a stalled angle of attack, $\alpha=35^\circ$. 
}
    \label{fig:FP_SPPOD}
\end{figure}

The first four modes of space-only POD and first two modes from IPhaB POD (shown at different phase instances along the vortex-shedding cycle) are shown in figures \ref{fig:FP_POD} and \ref{fig:FP_SPPOD} respectively. Space-only POD represents the alternating leading- and trailing-edge vortex shedding across a cascade of modes, with the first space-only mode (c.f., figure \ref{fig:FP_POD}(a)) being the time average of the flow field. The second and third mode (c.f., figures \ref{fig:FP_POD}(b and c)) show regions of vorticity that are up-down symmetric about $y=0$, with a matched downstream spatial wavelength. These modes are identical apart from a streamwise shift in vorticity to capture convection of the shed vortices. Modes 4 (c.f. figure \ref{fig:FP_POD}(d)) and 5 (not shown in this paper) from space-only POD have two vortex structures at each $x$ location at different vertical heights, representing harmonics of modes 2 and 3. For this problem, and generally for flows driven by a time-varying large scale, each space-only POD mode (or pair of modes) does not directly correspond to instantaneous flow structures. Each higher-index space-only POD mode inherits content from the large scale, which for more complex problems can be difficult to separate from information about other scales or processes  also represented by that mode. This mixing of large-scale content across modes is highlighted here for space-only POD, but is a general feature of decomposition techniques that require time-constant modes, such as standard implementations of dynamic mode decomposition and operator-based decomposition techniques (e.g., linear stability and resolvent analysis).

By comparison, the first IPhaB POD mode, shown in figures \ref{fig:FP_SPPOD}(a-d), preserves the leading- and trailing-edge vortex-shedding process that drives the physical system. Four instances of the second IPhaB POD modes, shown in figures \ref{fig:FP_SPPOD}(e-h), contain low intensity activity between the vortices identified in the first mode. The physical meaning and relevance of this and higher modes will be discussed later. %These low intensity regions are corrections to the first mode required by imperfect matching across the large-scale phase when building the data matrix. This outcome is a result of small temporal interpolation errors when matching the phase; subsequent plots will demonstrate that the energy associated with this and higher modes is very small.
The IPhaB POD mode is of size $n_s \cdot  n_n$, while the original data set is of size $n_s \cdot n_t$, where $n_n$ is again the number of samples in one period and $n_t$ is the total number of data samples. In general, $n_n$ is much smaller than $n_t$, representing a reduction in the size of the mode versus the original system. 

\begin{figure}
    \centering
    \includegraphics[width=\linewidth]{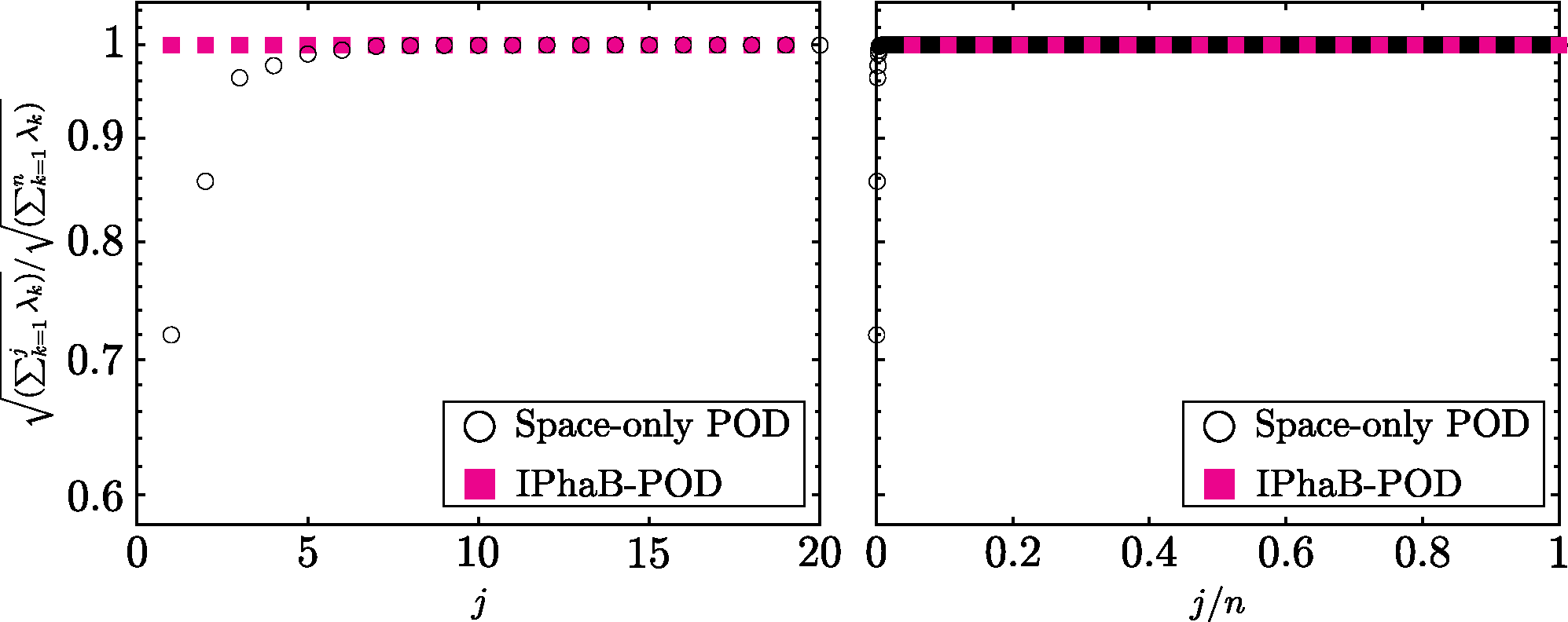}
    \caption{Cumulative energy versus mode index ($j$) for space-only POD and IPhaB POD. The x-axis in (b) is normalized by the total number of modes for the respective method (generically written as $n$). For this flat plate problem, there are a total of 19 IPhaB POD and 1628 space-only POD modes.}  %as defined as the sum of the square of the singular values normalized by their total squared sum (a) for all and (b) for the first 7.5\% of the flat plate modes. The ``energy'' is plotted against the fraction of total modes (there are drastically different numbers of space-only and IPhaB POD modes for the same amount of data.}
    \label{fig:FP_sigval}
\end{figure}

We show in figure \ref{fig:FP_sigval} the cumulative energy associated with the two POD representations, using the eigenvalues $\lambda_1, \lambda_2, \dots$ from the eigen-problem in (\ref{eqn:eval_disc}) (and its analog in space-only POD). The first IPhaB POD mode captures nearly all of the system energy while it takes the first 7 space-only POD modes to capture a similar amount of energy. This efficient representation of the large-scale vortex-shedding process is consistent with figures \ref{fig:FP_POD} and \ref{fig:FP_SPPOD}, in which multiple strong modes are observed for space-only POD and a single strong mode is observed for IPhaB POD. For IPhaB POD the energy associated with the second mode pictured in figures \ref{fig:FP_SPPOD}(e-h) (and subsequent ones) is substantially smaller than that of the first mode. This second mode (and higher ones) reflect slight corrections to the vortex-shedding captured by the first mode, required because of small numerical errors in the temporal interpolation procedure described in section \ref{sec:SPPOD-algorithm}. These small numerical errors produce a slight mismatch in the phase of the large scale across various cycles in building the data matrix (\ref{eqn:XSPPOD}). For this problem this interpolation error leads to modes providing small corrections in the interstices between the leading- and trailing-edge vortex process conveyed within the first mode, which have negligible energy associated with them.

%The first mode of a snapshot and space-phase POD decomposition are shown in figures \ref{fig:FP_POD}a and \ref{fig:FP_SPPOD}a respectively. The first modes of the two methods are qualitatively different. 
%The mode is a space-phase mode, and other phases of the first mode show the vortices convecting downstream. The large decay after the first IPhaB POD mode can be observed in the eigenvalues shown in \ref{fig:FP-E}, indicating the first mode contains almost all of the energy in the flow field. As this is simply a periodic limit cycle, a single periodic mode with a reported period of repetition is able to fully represent the flow with no additional modes required. 

We highlight two key caveats to this energy comparison. First, each IPhaB POD mode is of a larger dimension than a space-only POD mode ($n_s\cdot n_n$ versus $n_s$, respectively). To account for this difference in mode size, figure \ref{fig:FP_sigval}(b) shows that a smaller percentage of the total number of modes that each method produces is required for space-only POD modes compared to the IPhaB POD modes in order to represent a comparable amount of energy in the flow field. 

Second, it is perhaps natural to expect energy efficiency of the IPhaB POD representation relative to space-only POD for this canonical problem with a single large scale. For problems involving multiple scales, multiple modes (say, $k$) will typically be kept even in a truncated representation. In this setting, the large-scale behavior is typically the most energetic (by far) in the system. As such, better isolating the large-scale behavior to a single mode and distinct physical processes to other modes may not lead to a more energy-efficient representation than having the various scales mixed, energetically optimally, across the first $k$ modes. While this comparison of energy efficiency in representation will be explored for the shock problem below, we highlight that a comprehensive assessment of representation efficiency is subtle and should likely account for other factors beyond energy (both space-only POD and IPhaB POD are energetically optimal). This complexity of assessing model effectiveness is a  motivator for dynamic mode decomposition, which is an optimal (linear) model regression of the dynamics that might prioritize energetically small but dynamically important processes (see, for example, \citet{RowleyandDawson}).
Our aim in this work is to create an energetically optimal decomposition, retaining the beneficial properties of POD, that provides a more physically intuitive and interpretable representation for problems driven by a large scale, where this large scale and other behavior tethered to it are better isolated into modes more reflective of instantaneous structures. Integrating this representation into effective reduced order models, and quantifying model efficiency, is an area of future work.

\subsection{Cone-Cylinder at M=6}

We next perform IPhaB POD analysis on the cone-cylinder shock problem depicted in figure \ref{fig:overview_prob}(e-h), using the data set from \citet{Duvvuri}. To characterize the large-scale shock motion for this problem as described in section \ref{sec:large-scale-deep-dive}, a Canny edge detection method in MATLAB 2023b Images Processing tool box was used, with appropriate thresholds for a sparse identification of the shock-front. False shock-fronts identified by the algorithm were removed using a recursive moving median filter with window sizes from 3 to 40 time instances. With the locations from the sparse identification of the shock-front, we used a spline fit to form a continuous representation of the shock front. Due to the axisymmetric nature of the cone-cylinder body, all spatial positions below the axis of symmetry were labelled to be zero (evident in the filtered image in figure \ref{fig:shock_LS_identify}(b)). We smoothed the spatially integrated signal using a Savitzky Golay filter. We determined the period length and the start and end of each cycle based on the local maxima of $G$. The period length varied from $6.4912 \times 10^{-4}$  to $8.9474 \times 10^{-4}$ seconds with an average period length of $\bar{T}=7.8763 \times 10^{-4}$ seconds. Based on the thresholds $\epsilon^*$ and $\delta^*$ stated in section \ref{sec:large-scale-deep-dive} for this problem, 48 out of 124 cycles were determined to be near-periodic, with the nominal periodic cycle $\gamma^p$ satisfying $\delta^p = 0.0199$ and $\epsilon^p=0.2321$. In our comparison between IPhaB POD and space-only POD, space-only POD was built using all snapshots, and not just those from the near-periodic cycles, to be more representative of a typical implementation of that technique.

\begin{figure}
    \centering
    \includegraphics[width=\linewidth]{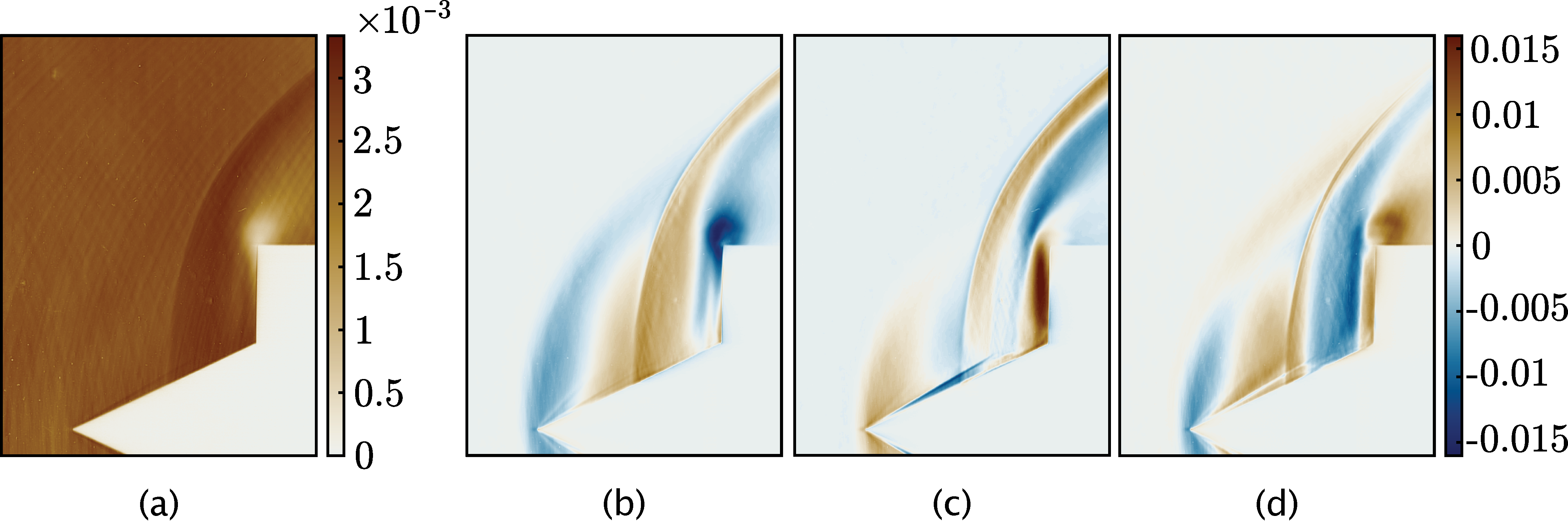}
    \caption{(a) First, (b) second, (c) third, and (d) fourth space-only POD modes for shock pulsations over a cone-cylinder body.}
    \label{fig:CC-POD}
\end{figure}

\begin{figure}
    \centering
    \includegraphics[width=0.9\linewidth]{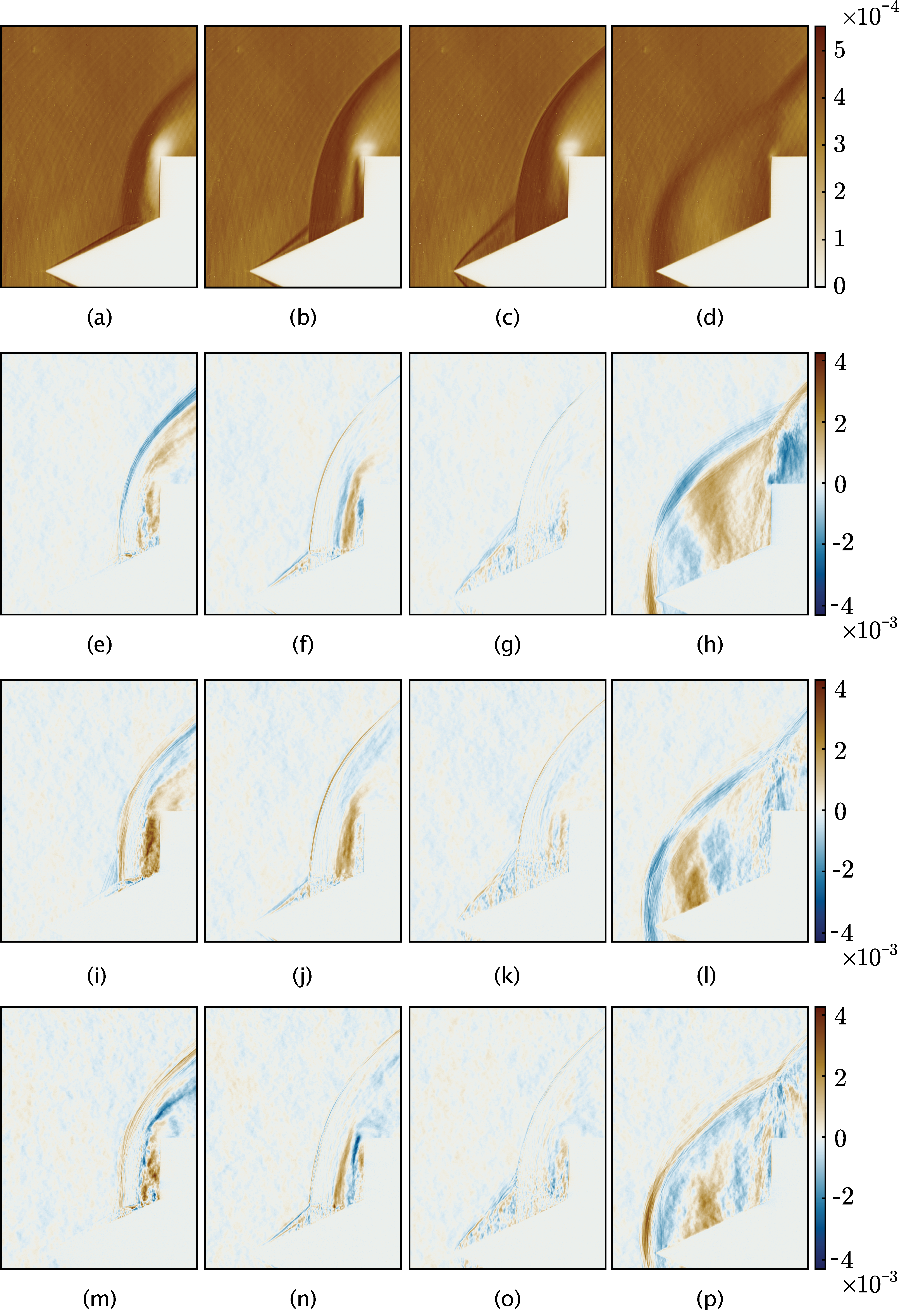}
    \caption{The (a-d) first, (e-h) second, (i-l) third, and (m-p) fourth IPHaB POD modes for shock pulsations over a cone-cylinder body. For each mode, four equally distributed phase instances (1st (a,e,i,m), 12th (b,f,j,n), 23rd (c,g,k,o), and 34th(d,h,l,p) out of 44 phases) are shown. }
    \label{fig:CC-SPPOD}
\end{figure}

The first four space-only and IPhaB POD modes are shown in figures \ref{fig:CC-POD} and \ref{fig:CC-SPPOD} respectively (with the IPhaB POD modes shown over four instances of a characteristic large-scale cycle). The first four space-only POD modes shown in figure \ref{fig:CC-POD} %all involve a shock-front at a shared spatial location. 
show features of the shock at multiple locations in the same mode. For example, mode 2 (c.f. figure \ref{fig:CC-POD}(b)), shows a shock front at the same location as mode 1, but also shows visually evident shock features at the tip of the cone, reflecting the behavior of the shock at a different part of the shock's undulation cycle. Mode 3 (c.f. figure \ref{fig:CC-POD}(c)) shows these two shock features as well as a third very near to and almost parallel with the cone, corresponding to yet another phase of the shock undulation cycle. The motion of the shock front is reconstructed through the summation of these modes (and subsequent ones, not pictured) through the canceling of the shock features in certain regions and enhancement in others as a function of the time-varying amplitude of each mode. The large-scale structure of the shock undulation is hard to deduce from these modes, as all the phases of its motion are mixed together across the modes. In addition, finer-scale physical processes summarized in section \ref{sec:data_overview} (and given in greater detail in \citep{Duvvuri}) such as the separation dynamics, triple point, and supersonic jet, are not visible, nor is it clear how those features are synchronized with the shock front dynamics.%. An indication of how these features evolve as the shock undulates is also not readily indicated by the modes.  

The first IPhaB POD mode (c.f. figures \ref{fig:CC-SPPOD}(a-d)) represents the time-varying motion of the shock front, which consists of a conical shock, separation shock, and a bow shock, over a single representative cycle. The shock front dynamics of the mode reflect the dynamics observed in the data, but are simpler than the original data, with fewer small-scale fluctuations observed. The shock front often appears as a sharp edge in the mode, but in some time instances appears more blurry and spatially distributed. Time instances where the shock is less sharp (for example, figure \ref{fig:CC-SPPOD}(d)) reflect that there are significant cycle-to-cycle variations in the spatial location of the shock in that part of the cycle, while time instances where the shock appears very sharp (for example, figures \ref{fig:CC-SPPOD}(b,c)) reflect that the shock positioning is very consistent across cycles. %The increased sharpness in representing the shock in figures \ref{fig:CC-SPPOD}(b)-(c), relative to figures \ref{fig:CC-SPPOD}(a) and (d), reflects the imperfect cycle periodicity inherent to this flow. 
Figures \ref{fig:CC-SPPOD}(b,c) are phases of the cycle where the
separation bubble grows under the adverse pressure gradient induced by the coalescence of the cone and cylinder shocks. We learn from the sharpness of the mode that this process is highly regular and repeatable across cycles. %This first-mode sharpness in (b)-(c) indicates smaller cycle-to-cycle variation than in plots (a) and (d), where the shock representation in the modes is blurrier. 
Figures \ref{fig:CC-SPPOD}(a,d), which show slightly blurrier shock fronts, %These plots 
correspond to the collapse (d) and reformation (a) of the separation bubble and subsequently of the shock system. We learn from the blurriness of the mode that these processes are less regular and more chaotic cycle-to-cycle.

\begin{figure}[h]
    \centering
    \includegraphics[width=\linewidth]{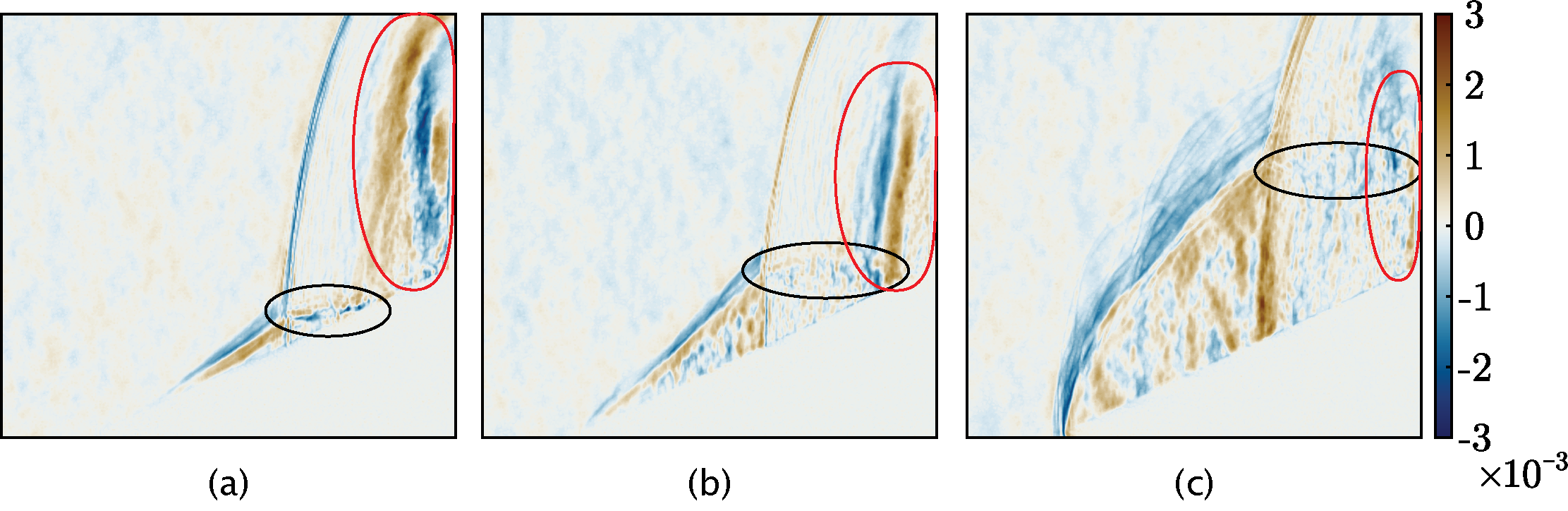}
    \caption{Three phase instances (18th (a), 28th (b), and 38th (c) of IPhaB POD Mode 2 zoomed in to highlight the evolution of the supersonic jet (black ellipses) and the activity near the base cylinder (red ellipses). }
    \label{fig:CC-SPPOD_2_Zoomed_in}
\end{figure}

The higher IPhaB POD modes contain details of smaller-scale physical processes, whose dynamics follow the cycle of the shock front. Small scale features, such as the supersonic jet and the triple point, are visible in the separation region, particularly in the middle phases of the cycle, including the 12th (c.f. figures \ref{fig:CC-SPPOD}(f, j, n)) and 23rd instances along the cycle (c.f. figure \ref{fig:CC-SPPOD}(g, k, o)). The movement of the supersonic jet and the triple point upward, away from the axis of symmetry can be seen in the first three instances shown of the second (e)-(g), third (i)-(k), and fourth (m)-(o) modes. Three more instances of the second IPhaB POD mode zoomed in near the cone-cylinder body are shown in figure \ref{fig:CC-SPPOD_2_Zoomed_in}. The supersonic jet is marked in black ellipses and the activity near the base cylinder is marked in red ellipses. As the supersonic jet moves away from the axis of symmetry, the strength of the jet also weakens, indicated by the decrease in the activity in the mode. The location of the strongest activity downstream of the shock front also changes at different phases of the shock pulsation. In the first phase instance (c.f. figure \ref{fig:CC-SPPOD_2_Zoomed_in}(a)), the strongest activity is present near the wall of the base cylinder and from the wall of the cone all the way to shock front. At the second phase instance (c.f. figure \ref{fig:CC-SPPOD_2_Zoomed_in}(b)), the activity primarily remains near the wall of the base cylinder, though the activity decreases above the base cylinder shoulder and the separation bubble associated with the boundary layer along the cone wall starts developing. In the last phase instance (c.f. figure \ref{fig:CC-SPPOD_2_Zoomed_in}(c)), the separation bubble is near its largest size and the supersonic jet becomes less coherent. The strong growth in bubble size from figure \ref{fig:CC-SPPOD_2_Zoomed_in}(b) to (c) demonstrates that the separation process is dominated by the final quarter of the undulation cycle.

While these distinct smaller-scale processes are visible across modes 2-4, it should also be noted that features of the first mode are also inherited in modes 2-4 near the early and late parts of the cycle. These artifacts are evident in, for example, figure \ref{fig:CC-SPPOD}(h) where large-wavelength flow structures dominate the mode's behavior. These phase instances where the large-scale shock is evident in higher modes correspond with the instances when the cycle-to-cycle variations in the shock motion are most pronounced. These cycle variations require that changes to large-scale shock features be conveyed by modes 2-4, in addition to the smaller-scale processes of primary interest. This imperfect isolation of the large and smaller scales is a possible issue to be addressed in future work, though we also highlight that this outcome provides information about instances when cycle-to-cycle variations are more and less pronounced.

\begin{figure}
    \centering
    \includegraphics[width=0.8\linewidth]{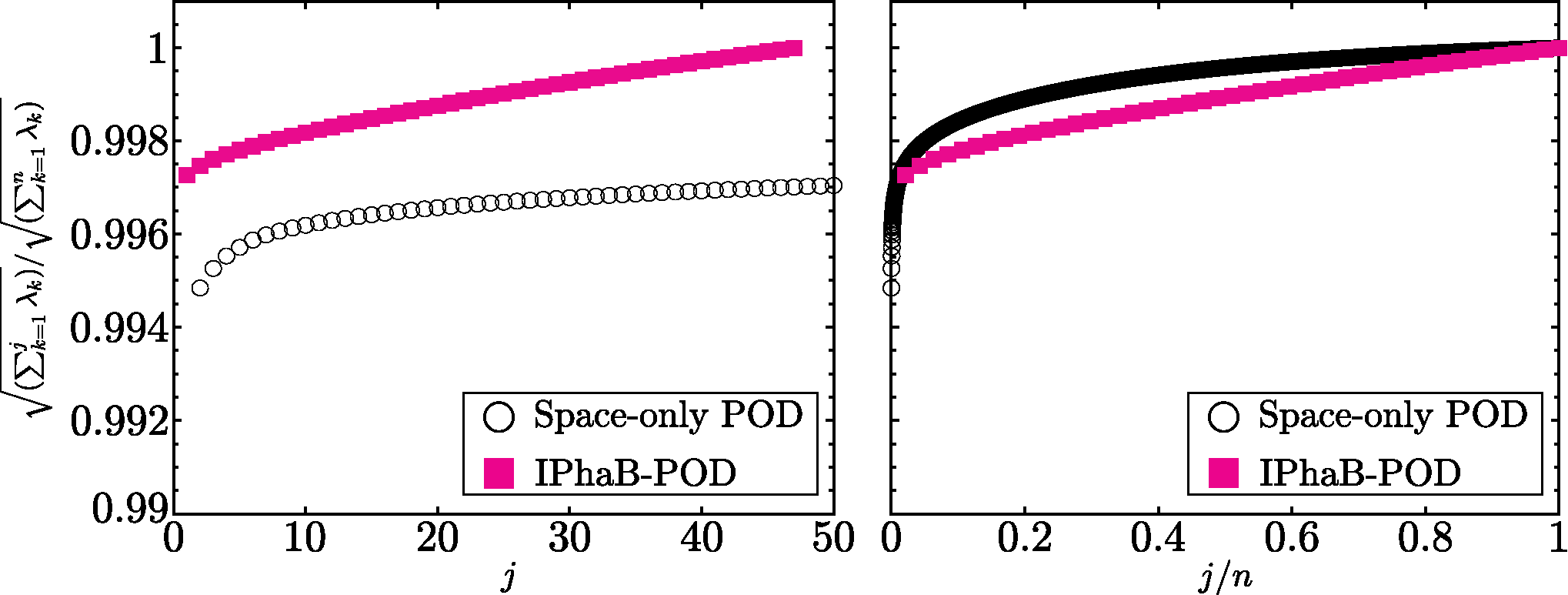}
    \caption{Cumulative ``energy'' as defined as the sum of the square of the singular values normalized by their total squared sum (a) for all and (b) for the first 10\% of the shock pulsation modes. The ``energy'' is plotted against the fraction of total modes (there are drastically different numbers of space-only and IPhaB POD modes for the same amount of data.}
    \label{fig:shock_sigval}
\end{figure}

To quantify and compare the efficiency of these decompositions, similar to the flat plate demonstration, we can evaluate the cumulative energy. We show the cumulative energy for the shock pulsations in figure \ref{fig:shock_sigval}. In figure \ref{fig:shock_sigval}(a), we show the cumulative energy plotted against the mode index. All of the IPhaB POD modes shown and fifty of the 5601 space-only POD modes are shown. The first mode of IPhaB POD has a larger percentage of the energy than the first mode of space-only POD. The first modes of both methods contain over 99\% of the energy of the data. This high energy representation is likely due to the nonzero pixel intensity across the vast majority of any given raw image. These nonzero values were not removed to compute the modes, since to facilitate a clear comparison across IPhaB and space-only POD the mean was not subtracted from the data prior to analysis. %We further conclude the first space-only POD mode is the temporal mean while the first IPhaB POD is the ``phase locked" mean, where the phase is locked based on the shock pulsation. %We also observe IPhaB POD modes reach 100\% of the energy much faster compared to space-only POD modes however it is important to consider the size of each IPhaB POD mode is much larger than each space-only POD mode. 
In figure \ref{fig:shock_sigval}(b), we plot the cumulative energy against the mode index normalized with the total number of modes that each method produces. Once normalized, we observe that it takes a smaller percentage of the total space-only POD modes compared to the IPhaB POD modes to contain the same amount of energy in the flow field, just as it had with the low Reynolds number flat plate problem. 

%A proper integration of IPhaB POD into reduced order models is left to future work, as is a comprehensive representation of model efficiency, which would likely account for efficiency in both conveying energy and dynamics of the overall system. In this article, we focus on developing an IPhaB POD method that provides modes that more intuitively represent instantaneous flow structures. We demonstrated that the method also provides additional clarity to the oscillatory behavior of the large-scale shock, and to how the finer-scale process associated with the triple point motion, supersonic jet, and separation bubble are tethered to the shock's behavior.

\section{Conclusion}

A proper orthogonal decomposition framework was developed, IPhaB POD, that is appropriate for systems driven by an inherent large-scale  with strongly (but imperfectly) periodic content.  
For these systems, the proposed method leverages a dynamical systems representation of a set of restricted dynamics sampled from the full dynamical state. These restricted dynamics are provided by the user and expected to convey information about the various cycles of the large-scale behavior. For example, for the flat plate problem considered in this article, the restricted dynamics were of the lift coefficient, which clearly distinguished the various phases of the large-scale vortex shedding process. We provided a framework that uses these restricted dynamics to identify which cycles of the full system dynamics are associated with a characteristic large-scale orbit (as opposed to those corresponding to cases where the large-scale undergoes an extreme trajectory). All cycles that contained a characteristic large-scale trajectory were collected into a set $\Gamma$ and were considered for the POD analysis; others were discarded. We also used this framework to define the cycle from the data that contains a most typical large-scale trajectory, $\gamma^p$.

An averaging operation and inner product were then defined that referred each instance from any one of the trajectories in $\Gamma$ to the nearest corresponding instance in $\gamma^p$. This process allowed for the various cycles within $\Gamma$, which each contained meaningfully similar large-scale dynamics but with cycle-to-cycle differences, to be compared. Once the inner product and expectation were defined that allowed an appropriate comparison of the various near-periodic trajectories in $\Gamma$, a typical POD was performed using the standard eigenvalue problem of a suitably constructed data matrix. The resulting decomposition inherits the desirable properties of energetic optimality and mode coherence of POD.

The method was tested on two test problems. The first problem was from a high-fidelity computation of a canonical $Re = 100$ flow past an airfoil at the large angle of attack of $\alpha=35^\circ$. In this laminar flow, the only behavior is of periodic vortex shedding that alternates from the leading and trailing edges. IPhaB POD captured this vortex shedding with a single time-varying mode. This was contrasted with space-only POD, which required %many modes, both harmonics to capture a lack of up-down symmetry in the original flow field and phase-shifted modes to capture convection.
a cascade of modes. The modes came in pairs to capture the convective nature of the vortex-shedding process, and the modes were sequenced so that higher modes were of higher spatial frequency to compensate for the incorrect up-down symmetry imposed by each mode pair. 

The second problem was an experimental data set from schlieren data of high-Reynolds and Mach number flow past a cone-cylinder system. IPhaB POD largely isolated the large-scale shock oscillations within the first time-varying mode, and yielded clearly identifiable small-scale activity in the higher time-varying modes. These higher modes represented the boundary layer separation as well as the motion of the triple point and its associated supersonic jet that are known to be key processes in this physical system \citep{Duvvuri}. These higher modes moreover evolved with the shock motion, providing more physically intuitive representations of instantaneous flow features for this complex system. This decomposition was contrasted with space-only POD, which contained significant features of the large-scale shock across all depicted modes (and ensuing ones not shown), obscuring any visible evidence of the smaller-scale dynamics key to sustaining the near-periodic process.  %showed how the small-scale activity varied as a function of the large-scale shock pulsation phase and showed dynamics that were not visible in space-only POD modes of the same flow field. 

%While the method yielded modes that were more emblematic of instantaneous flow structures, t
The energy efficiency of representing the dynamics via IPhaB POD was demonstrated to be comparable to space-only POD for the shock problem. This result is perhaps not surprising---both decompositions are energetically optimal with respect to their notion of energy, and it is likely that the additional processes depicted by IPhaB POD, associated with the separation bubble growth and triple point dynamics, are energetically small but dynamically significant. Future work could therefore expand an evaluation of representation efficiency to include dynamic (and not just energetic) considerations. It was also noted that for the shock problem, where the large-scale dynamics were not perfectly periodic, artifacts of the large scale were inherited in higher modes (though to a far smaller extent than for space-only POD). Techniques to address this leakage of large-scale behavior into higher modes is another possible avenue of future work.

\section*{Acknowledgements}
\noindent We would like to acknowledge Professor Subrahmanayam Duvvuri at the Indian Institute of Science and his graduate student Vaisakh Sasidharan for providing us with the cone-cylinder schlieren data used for analysis.

\section*{Funding}
\noindent This work was funded by the National Science Foundation under grant 2118209.

%Bibliography
%\bibliographystyle{unsrt}  
\bibliography{references}  
\bibliographystyle{plainnat}

\end{document}